\documentclass[aps,twocolumn,showpacs,preprintnumbers,prb,amsmath,amssymb,amsfonts,superscriptaddress,floatfix]{revtex4-1}
\usepackage{dcolumn}

\usepackage{graphics}
\usepackage{color}
\usepackage{amssymb}
\usepackage{amsmath}
\usepackage{overpic}
\usepackage{epstopdf}
\usepackage{bm}
\usepackage{graphicx}
\usepackage{subfigure}
\usepackage[latin1]{inputenc}

\begin{document}

\title{Electronic fitness function for screening semiconductors as thermoelectric materials}

\author{Guangzong Xing}
\affiliation{College of Materials Science and Engineering and Key Laboratory of Automobile Materials of MOE, Jilin University, Changchun 130012, China}
\affiliation{Department of Physics and Astronomy, University of Missouri-Columbia, Columbia, MO 65211, USA}
\author{Jifeng Sun}
\email{sunjif@missouri.edu}
\affiliation{Department of Physics and Astronomy, University of Missouri-Columbia, Columbia, MO 65211, USA}
\author{Yuwei Li}
\email{liyuw@missouri.edu}
\affiliation{Department of Physics and Astronomy, University of Missouri-Columbia, Columbia, MO 65211, USA}
\author{Xiaofeng Fan}
\affiliation{College of Materials Science and Engineering and Key Laboratory of Automobile Materials of MOE, Jilin University, Changchun 130012, China}
\author{Weitao Zheng}
\affiliation{College of Materials Science and Engineering and Key Laboratory of Automobile Materials of MOE, Jilin University, Changchun 130012, China}
\author{David J. Singh}
\email{singhdj@missouri.edu}
\affiliation{Department of Physics and Astronomy, University of Missouri-Columbia, Columbia, MO 65211, USA}

\date{\today}

\begin{abstract}
We introduce a simple but efficient electronic fitness function (EFF) that describes the electronic aspect of the thermoelectric performance.
This EFF finds materials that overcome the inverse relationship between
$\sigma$ and $S$ based on the complexity of the electronic structures
regardless of specific origin (e.g., isosurface corrugation,
valley degeneracy, heavy-light bands mixture,
valley anisotropy or reduced dimensionality).
This function is well suited for application in high throughput screening.
We applied this function to 75 different thermoelectric and
potential thermoelectric materials including full- and half-Heuslers,
binary semiconductors and Zintl phases.
We find an efficient screening using this transport function. The EFF
identifies known high performance $p$- and $n$-type Zintl phases and
half-Heuslers. In addition, we find some
previously unstudied phases with superior EFF.
\end{abstract}

\maketitle

\section{Introduction}

Direct thermal-to-electrical energy conversion
has made thermoelectric (TE) materials a current interest in energy technology.\cite{r1,r2,r3,r4,r5} TE performance is governed by the dimensionless figure of merit, $ZT$ = ($S^{2}\sigma T$)/$\kappa$, of the materials, where $S$, $\sigma$, $T$, $\kappa$ are the Seebeck coefficient, the electrical conductivity, the temperature, and the thermal conductivity, respectively. Efforts to increase $ZT$ have mainly focussed on the maximization of the power factor (PF=$S^{2}\sigma$) through optimal doping and band engineering,\cite{r6,r7,r8} and the reduction of $\kappa_{l}$ (the lattice part of $\kappa$).\cite{r9,r10} A key challenge is that high $ZT$ is a contraindicated property in the sense that the ingredients in $ZT$ show inverse relationships. Standard models of semiconductors such as the isotropic single parabolic band model do not lead to high $ZT$. For high power factor one needs high $S$, which can be obtained from high effective mass and low carrier density but oppositely for high $\sigma$. Good TE materials generally have complex electronic structures not characterized by a simple parabolic band. Thus the conflict between $\sigma$ and $S$ can be resolved. The challenge that we address here is how to efficiently identify materials with such a characteristic.

\begin{figure*}[t]
\centering
\includegraphics*[height=12cm,keepaspectratio]{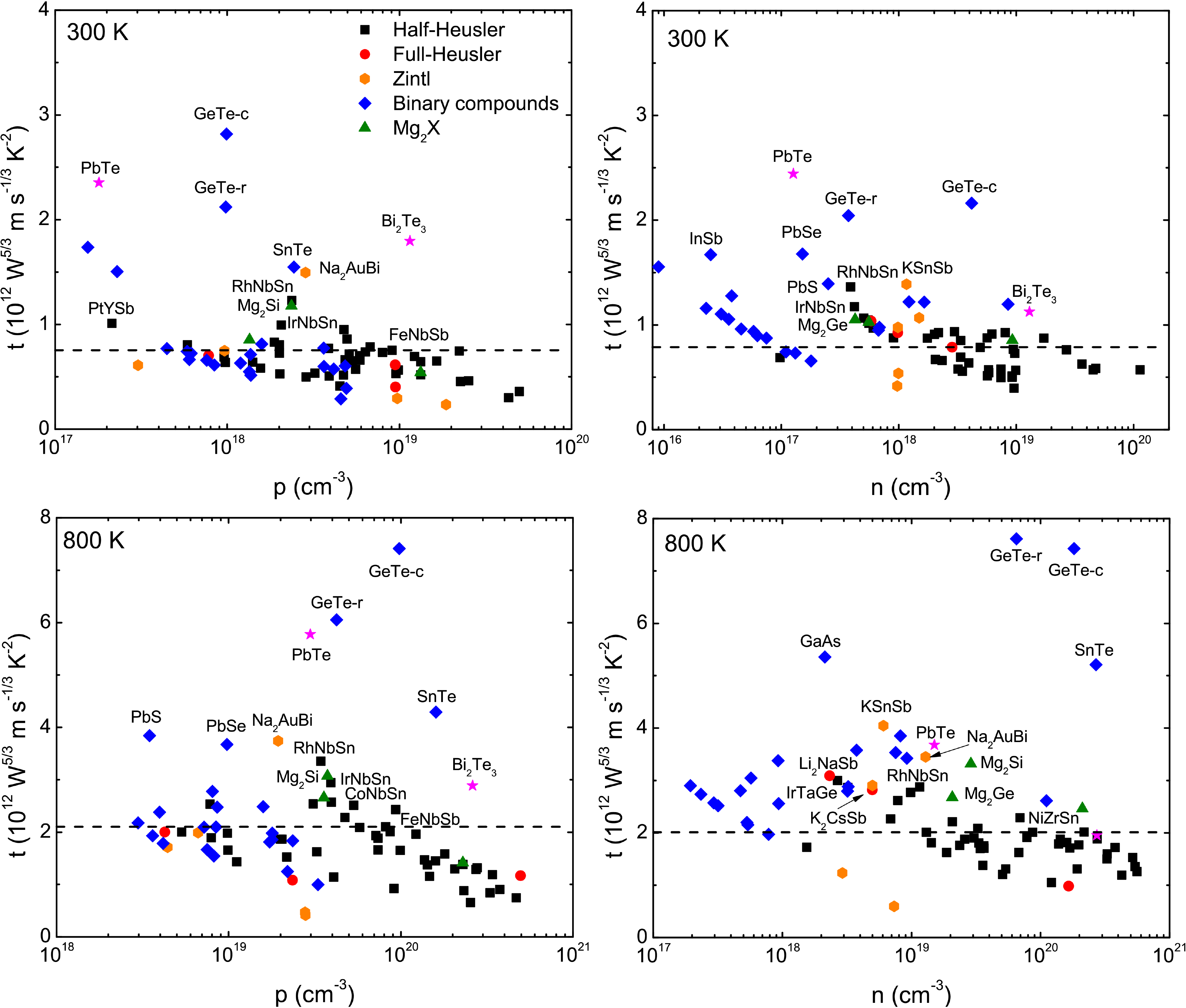}
\caption{\label{fig1} The optimum EFF ($t$) and corresponding carrier concentrations of all the compounds studied. The doted lines denote are criteria based good known TE materials (see text). Materials below this are less interesting
for thermoelectricity.}
\end{figure*}

\begin{figure*}[t]
\centering
\includegraphics*[height=12cm,keepaspectratio]{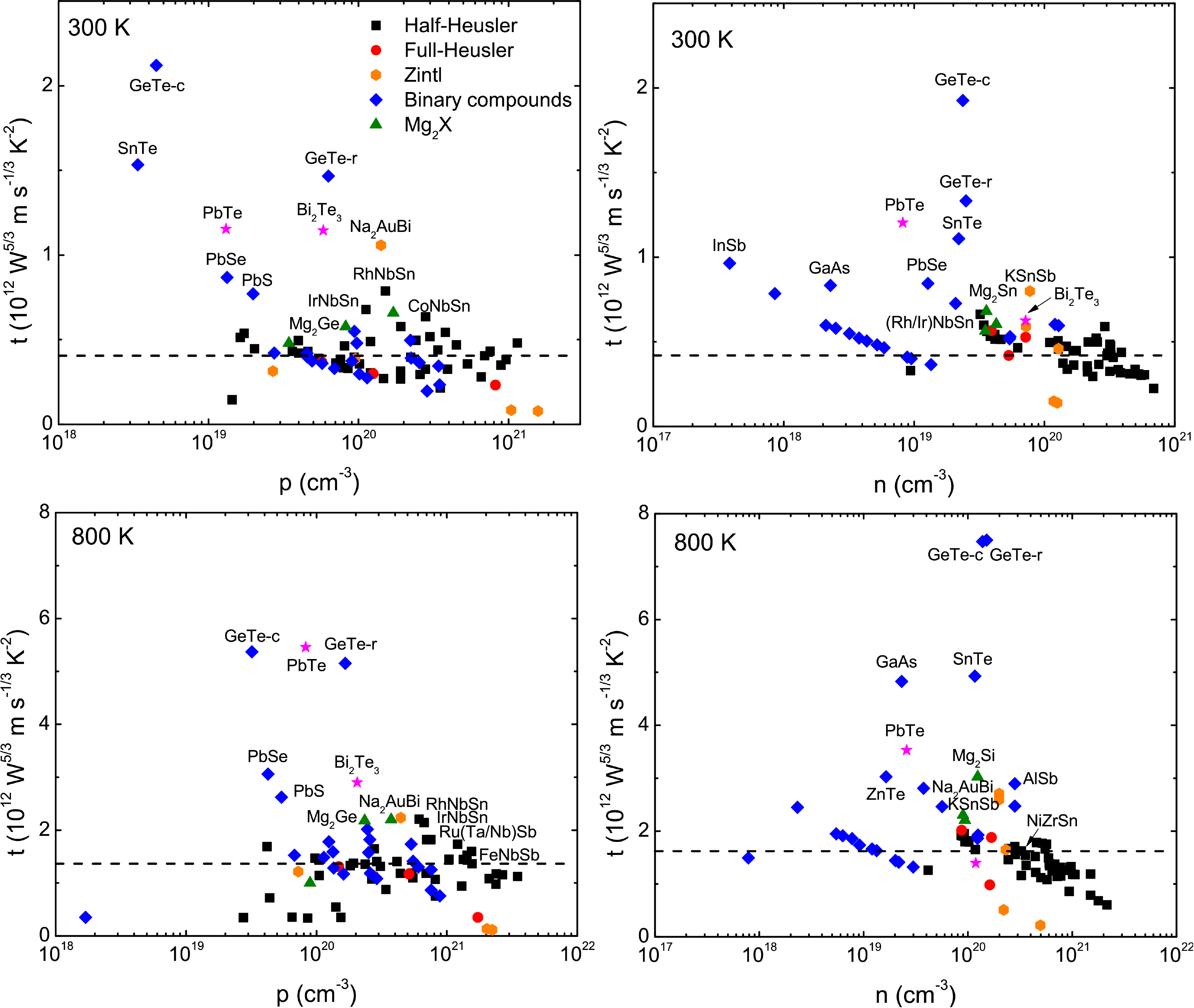}
\caption{\label{fig2} EFF ($t$) at the constant energy level of 0.05 eV below valence band maximum ($p$-type) and above conduction band minimum ($n$-type). The doted lines denote a screening criteria based on several known TE materials (see text). }
\end{figure*}

Here we present and explore the use of a transport function,
the Electronic Fitness Function (EFF) $t$ = ($\sigma$/$\tau$)$S^{2}$/$N^{2/3}$
that measures the extent to which a general complex band structure decouples $\sigma$ and $S$. Here $\tau$ is an inverse scattering rate. $\sigma$/$\tau$ can be obtained directly from band structure, but $\sigma$ and $\tau$ separately require detailed knowledge of the scattering. The EFF is low in isotropic parabolic band systems. This EFF can be directly evaluated based on the first-principles electronic structures and Boltzmann transport theory under the constant relaxation time approximation.\cite{r16,r17} It shows promise in relation to other measures for oxides.\cite{r16}
In the EFF, $N$ is the volumetric density of states, which is proportional to the density of states effective mass and Fermi energy as $N \sim (m^{\ast}_{dos})^{3/2}E_{F}^{1/2}$ for a parabolic band, where $E_F$ is relative to the band edge. The complexity of the electronic structure may come from multi-valley carrier pockets,\cite{vr1} valley anisotropy,\cite{isoene2,r17} band convergence,\cite{isoene3} heavy-light band combination,\cite{vr2,vr3} complex iso-energy surfaces,\cite{r16} reduced dimensionality,\cite{n1} and nonparabolic bands,\cite{add6,add7} all of which have been proved to be favorable for high TE performance in certain cases. With such a variety of favorable electronic structures, how does one efficiently find materials that are favorable? 

The proposed EFF is designed as such a universal function that incorporates all these features. Importantly, the 2/3 power of the density of states reduces the tendency of screens based on the calculated $S$ to find heavy mass semiconductors, which often do not conduct. This is important because heavy mass by itself is not sufficient to get high $ZT$.\cite{r16} Moreover, all the quantities in $t$ can be readily obtained from the band structure.
This is an important requirement for implementing an efficient
high throughput screening.

Here we investigate various ways of using the EFF
to identify potential TE materials. We use a set of 75 semiconducting materials,
 including half-Heuslers and full-Heuslers, binary semiconductors, and Zintl phases. We find that this transport function can efficiently screen the materials. We also identified some novel $p$- and $n$-type phases that exhibit favorable complex band structures in relation to the known TE materials.  

\section{Computational Methods}

\subsection{\textbf{Electronic structure calculations}}

The electronic structures were obtained with the all-electron general potential linearized augmented planewave (LAPW) method,\cite{r18} as implemented in the WIEN2k code.\cite{r19} Experimental lattice constants were used, while the atomic coordinates were relaxed when needed by total energy minimization using the Perdew, Burke and Ernzerhof (PBE) functional.\cite{r20} Then we employed the modified Becke-Johnson (mBJ) potential \cite{r21} for the electronic structure calculations with the relaxed structural parameters. This potential yields improved band gaps relative to standard GGA and LDA functionals. The LAPW sphere radii were chosen in the standard way. We did all calculations relativistically including spin-orbit except for the structure relaxations, where relativity for the valence states was treated at the scalar level.
We used the accurate LAPW plus local orbital method rather than the faster,
but sometimes less accurate APW+lo method. \cite{lapw+lo,apw+lo}
We used the highly converged choice, $R_{min}K_{max}$ = 9
for the planewave cutoff plus local orbitals for semicore states.
For the transport calculation we used at least 50000 k-points in the full
Brillouin zone for simple compounds with five or fewer atoms per cell,
and correspondingly dense meshes for more complex materials (note that
the size of the zone to sample decreases with the number of atoms in the cell).
For Bi$_2$Te$_3$, the local density approximation
was used due to the known deficiency of the
PBE functional compared to LDA for the structure of this
van der Waals compound.

\subsection{\textbf{Boltzmann transport calculations}}

The transport coefficients were obtained using the Boltzmann transport theory in the relaxation time approximation as implemented in the BoltzTraP code.\cite{r22} Within this relaxation time approximation, the electrical conductivity and Seebeck coefficient can be written as:

\begin{equation}
S(T,E_{F})=-\frac{1}{eTV}\frac{\int\sigma_{\alpha\beta}(E)(E-E_{F})f^{\prime}(T,E-E_{F})dE}{\int\sigma_{\alpha\beta}(E)f^{\prime}(T,E-E_{F})dE}
\label{eq1}
\end{equation}
and

\begin{equation}
\sigma(T,E_{F})=-\frac{1}{V}\int\sigma_{\alpha\beta}(E)f^{\prime}(T,E-E_{F})dE,
\label{eq2}
\end{equation}

\noindent where $V$ is the unit cell volume, $f^{\prime}$ is the energy derivative of the Fermi function at temperature T, and $\sigma_{\alpha\beta}(E)$ is the energy dependent transport function defined as:

\begin{equation}
\sigma_{\alpha\beta}(E)=e^{2}\int v_{\alpha}(k)v_{\beta}(k)\tau(k)\delta(E-E(k))d^{3}k,
\label{eq3}
\end{equation}

\noindent where $E(k)$ is the band energy and $v$ = $\nabla_{k}E$/$\hbar$ is the group velocity of carriers that can be directly derived from band structures. The energy dependent relaxation time $\tau$ can be very difficult to determine.
However, in the constant scattering time approximation (CSTA),
which assumes the energy dependence of the scattering rate
is negligible compared with the energy dependence of the electronic structure,
$\tau$ cancels in the expression for $S$.
The CSTA has been successfully applied in calculating Seebeck coefficients
for various TE materials.\cite{r11,r22,r23,r25,r26,r27,add1}
However, $\tau$ is still needed for $\sigma$ and the PF.
Various strategies have been applied for this.
The most common is to use an universal $\tau$ with a fixed value (e.g., 10$^{-14}$ s).\cite{r14,r151,r27} This misses the increased scattering caused by phonons at high $T$ as well as the doping dependence.\cite{r151} Furthermore, this is a very optimistic scenario since normally one expects scattering to increase both as $T$ is increased and as the carrier concentration is raised. The proposed EFF, $t$, through the $N^{2/3}$ factor is more conservative.
Importantly, relative to using using $\sigma S^2$/$\tau$, the EFF penalizes
heavy effective mass (note that heavy mass typically leads to low
$\sigma$).
It also penalizes high temperature as we use it for the numerator of $ZT$ including the factor of $T$
(using $t$ as an indicator of the numerator of $ZT$ amounts to using $\tau^{-1}\sim TN^{2/3}$).
In the following we present results based on $t$,
which is a function of both doping and temperature and
explore different ways of using this function to identify promising compounds.

\begin{center}
\begin{table*}
\caption{\label{tab:table1} Potential promising TE candidates with maximum and isoenergy electronic fitness function larger than the criterion compounds (see text) at 300 and 800 K for both $p$- and $n$-type materials. Note that the promising materials estimated from the isoenergy $t$ function include more compounds than the maximum $t$ function except for $n$-type binary compounds.} 
\resizebox{18cm}{!}{
\begin{tabular}{lcccclccc|lcccclccccccc}
\hline
\hline 
300 K& & & & & & & & & 800 K \\ 
$p$-type             &              &             &           &  &  $n$-type  &         &             &           &   
$p$-type             &              &             &           &  &  $n$-type  &         &             &           &\\
Mater.    & $t_{max}$($p$)  & Mater. & $t_{0.05}$($p$) &    & Mater.  & $t_{max}$($n$)  & Mater.   & $t_{0.05}$($n$) & 
Mater.    & $t_{max}$($p$)  & Mater. & $t_{0.05}$($p$) &    & Mater.  & $t_{max}$($n$)  & Mater.   & $t_{0.05}$($n$) & \\
\hline 
 GeTe-c& 2.82 &GeTe-c& 2.12 &  & PbTe & 2.44   &  GeTe-c & 1.93 & 
 GeTe-c& 7.41 & PbTe & 5.46 &  & GeTe-r & 7.61 &  GeTe-r & 7.50 &         \\  
 PbTe  & 2.35 & SnTe & 1.53 &  & GeTe-c & 2.16 &  GeTe-r & 1.33 & 
 GeTe-r& 6.05 &GeTe-c& 5.37 &  & GeTe-c & 7.43 &  GeTe-c & 7.47 &         \\
 GeTe-r& 2.12 & GeTe-r & 1.47 &  & GeTe-r & 2.04 &  PbTe & 1.20 & 
 PbTe  & 5.78 & GeTe-r & 5.15 &  & GaAs & 5.36 &  SnTe & 4.93 &         \\
 Bi$_{2}$Te$_{3}$  & 1.80 & PbTe & 1.15 &  & PbSe & 1.68 &  SnTe & 1.11 & 
 SnTe  & 4.29 & PbSe & 3.06 &  & SnTe & 5.21 &  GaAs & 4.83 &         \\
 PbSe  & 1.74 & Bi$_{2}$Te$_{3}$ & 1.14 &  & InSb & 1.67 &  InSb & 0.96 & 
 PbS   & 3.84 & Bi$_{2}$Te$_{3}$ & 2.84 &  & AlSb & 3.85 &  PbTe & 3.52 &         \\
 SnTe  & 1.55 & PbSe & 0.87 &  & InAs & 1.55 &  PbSe & 0.85 & 
 PbSe  & 3.67 & PbS  & 2.62 &  & PbTe & 3.68 &  ZnTe & 3.02 &         \\    
 PbS   & 1.50 & PbS  & 0.77 &  & PbS  & 1.39 &  GaAs & 0.83 & 
 Mg$_{2}$Si & 3.07 & Mg$_{2}$Si  & 2.20 &  & PbS & 3.58 &  Mg$_{2}$Si & 3.02 &         \\
 Mg$_{2}$Si & 1.18 & Mg$_{2}$Si & 0.66 &  & GaAs & 1.28&  InAs & 0.79 & 
 Bi$_{2}$Te$_{3}$ & 2.89 & Mg$_{2}$Ge & 2.18 &  & GaP & 3.53 &  AlSb & 2.89 &         \\
 Mg$_{2}$Ge & 0.85 & Mg$_{2}$Ge & 0.58 &  & AlSb & 1.22 &  PbS & 0.73 & 
 GaP   & 2.78 & GaP & 2.01 &  & PbSe & 3.42 &  PbSe & 2.81 &         \\
 GaP   & 0.81 & GaP & 0.55 &  & GaP & 1.22 &  Mg$_{2}$Sn & 0.68 & 
 Mg$_{2}$Ge & 2.66 & InP & 1.82 &  & ZnTe & 3.38 &  GaP & 2.47 &         \\  
 AlP   & 0.77 & AlP & 0.50 &  & SnTe & 1.20 &  Bi$_{2}$Te$_{3}$ & 0.61 & 
 AlP   & 2.48 & GaAs & 1.78 &  & Mg$_{2}$Si & 3.32 &  PbS & 2.46 &         \\
 InSb  & 0.77 & InP & 0.48 &  & InN-c & 1.16 &  Mg$_{2}$Si & 0.60 & 
 InP   & 2.48 & AlP & 1.73 &  & InAs & 3.04 &  InAs & 2.45 &         \\    
 /  & / & Mg$_{2}$Sn & 0.48 &  & Bi$_{2}$Te$_{3}$ & 1.13 &  AlSb & 0.60 & 
 GaAs  & 2.38 & InAs & 1.59 &  & InP & 2.90 &  Mg$_{2}$Ge & 2.30 &         \\  
 /  & / & InSb & 0.42 &  & InN-h & 1.10 &  InN-c & 0.60 & 
 AlSb  & 2.17 & AlAs & 1.58 &  & AlAs & 2.88 &  Mg$_{2}$Sn & 2.21 &         \\
 /  & / & GaAs & 0.42 &  & InP & 1.05 &  GaP & 0.60 & 
 AlAs  & 2.10 & InSb  & 1.53 &  & InN-h & 2.81 &  InN-c & 1.95 &         \\    
 /  & / & / & / &  & Mg$_{2}$Ge & 1.05 &  InN-h & 0.58 & 
 /  & / & AlSb & 1.48 &  & AlP & 2.79 &  AlAs & 1.93 &         \\
 /  & / & / & / &  & Mg$_{2}$Si & 1.03 &  Mg$_{2}$Ge & 0.56 & 
 /  & / & ZnS   & 1.41 &  & CdTe  & 2.73 &  InN-h & 1.91 &         \\  
 /  & / & / & / &  & AlAs & 0.98 &  InP & 0.55 & 
 /  & / & / & / &  & Mg$_{2}$Ge & 2.68 &  AlP & 1.86 &         \\
 /  & / & / & / &  & CdTe  & 0.96 & AlAs & 0.53 & 
 /  & / & / & / &  & InSb & 2.61 &  InP & 1.86 &         \\    
 /  & / & / & / &  & AlP & 0.95 &  CdTe & 0.52 & 
 /  & / & / & / &  & CdSe & 2.57 &  CdTe & 1.73 &         \\  
 /  & / & / & / &  & ZnTe & 0.94 &  AlP  & 0.52 & 
 /  & / & / & / &  & InN-c & 2.56 &  CdSe & 1.66 &         \\
 /  & / & / & / &  & CdSe & 0.90  &  ZnTe  & 0.51 & 
 /  & / & / & / &  & ZnSe  & 2.52 &  ZnSe & 1.63 &         \\    
 /  & / & / & / &  & ZnSe & 0.87  &  CdSe & 0.48 & 
 /  & / & / & / &  & Mg$_{2}$Sn & 2.46 &  / & / &         \\     
 /  & / & / & / &  & Mg$_{2}$Sn & 0.85  & ZnSe & 0.47 & 
 /  & / & / & / &  & CdS & 2.20 &  / & / &         \\
 /  & / & / & / &  & / & /  &  / & / & 
 /  & / & / & / &  & ZnS & 2.15 &  / & / &         \\        
 Na$_{2}$AuBi & 1.49 & Na$_{2}$AuBi & 1.06 &  & KSnSb & 1.39 &  KSnSb & 0.80 & 
 Na$_{2}$AuBi & 3.74 & Na$_{2}$AuBi & 2.23 &  & KSnSb & 4.05 &  Na$_{2}$AuBi & 2.71 &         \\ 
 RhNbSn  & 1.23 & RhNbSn & 0.79 &  & RhNbSn & 1.36 &  RhNbSn & 0.66 & 
 RhNbSn  & 3.35 & RhNbSn & 2.21 &  & Na$_{2}$AuBi & 3.45 &  KSnSb & 2.59 &         \\
 PtYSb   & 1.01 & IrNbSn & 0.68 &  & IrNbSn & 1.18 &  IrNbSn & 0.60 & 
 IrNbSn  & 2.94 & IrNbSn & 2.14 &  & Li$_{2}$NaSb & 3.09 &  Li$_{2}$NaSb & 2.02 &         \\ 
 IrNbSn  & 1.00 & CoNbSn & 0.64 &  & LiAsS$_{2}$ & 1.07 &  RuVSb & 0.59 & 
 IrTaGe  & 2.57 & RuTaSb & 1.82 &  & IrTaGe & 3.00 &  RhNbSn & 1.95 &         \\  
 CoNbSn  & 0.95 & CoHfSb & 0.58 &  & IrTaGe & 1.06 &  Na$_{2}$AuBi & 0.59 & 
 RuTaSb  & 2.54 & RuNbSb & 1.82 &  & LiAsS$_{2}$ & 2.90 &  IrTaGe & 1.92 &         \\ 
 RuNbSb  & 0.86 & IrTaGe & 0.54 &  & IrTaSn & 1.05 &  Li$_{2}$NaSb & 0.57 & 
 RhHfSb  & 2.53 & IrTaGe & 1.74 &  & RhNbSn & 2.88 &  IrNbSn & 1.91 &         \\ 
 RhHfSb  & 0.83 & PdYSb  & 0.54 &  & Li$_{2}$NaSb & 1.04 &  IrTaGe & 0.56 & 
 CoNbSn  & 2.51 & PtYSb  & 1.69 &  & IrTaSn & 2.87 &K$_{2}$CsSb & 1.88 &         \\ 
 NiYSb   & 0.81 & RuNbSb & 0.52 &  & RuNbSb & 1.02 &  IrTaSn & 0.54 & 
 RuNbSb  & 2.43 & RhHfSb & 1.65 &  & K$_{2}$CsSb & 2.82 &  RuTaSb & 1.82 &         \\ 
 PtScSb  & 0.81 & PtYSb  & 0.51 &  & Na$_{2}$AuBi & 0.98 &  RuNbSb & 0.53 & 
 IrTaSn  & 2.28 & CoNbSn & 1.60 &  & IrNbSn & 2.77 &  RuNbSb & 1.80 &         \\
 IrTaGe  & 0.79 & RuTaSb & 0.50 &  & FeNbSb & 0.97 &  K$_{2}$CsSb & 0.53 & 
 \textbf{FeNbSb} & 2.10 & IrTaSn & 1.52 &  & RuTaSb & 2.62 & IrTaSn & 1.79 &   \\
 PdHfSn  & 0.77 & NiYSb  & 0.50 &  & RuTaSb & 0.97 &  PdYSb  & 0.52 & 
        /&     /& PtScSb & 1.47 &  & PtYSb  & 2.29 &  PdYSb & 1.78 &         \\        
 RuTaSb  & 0.77 & RhHfSb & 0.49 &  & NiHfSn & 0.94  & RuTaSb & 0.52 & 
        /&     /& NiYSb  & 1.45 &  & IrZrSb & 2.27 & PtYSb  & 1.77 &             \\        
 \textbf{FeNbSb}& 0.75   & CoTaSn  & 0.48 &  &PdLaBi & 0.93 & FeNbSb & 0.52 &
        /&     /& RuVSb  & 1.45 &  & FeNbSb & 2.21 & NiYSb & 1.75 &                         \\              
        /&     /& IrTaSn & 0.47 &  & PdHfSn & 0.93 &  NiHfSn & 0.51 & 
        /&     /& CoHfSb & 1.44 &  & RuNbSb & 2.09 &  NiHfSn & 1.71 &         \\ 
 /& / & NiScSb & 0.46 &  & K$_{2}$CsSb & 0.93 &  PtYSb & 0.50 & 
 /& / & PdHfSn & 1.41 &  & NiYSb   & 2.01   & PdHfSn & 1.67 &         \\ 
 / & /& PdLaBi & 0.45 &  & NiYSb & 0.91 &  PdYBi & 0.50 & 
 /  & /& RhZrSb & 1.41 &  & PtHfSn & 2.01 &  NiYBi & 1.66 &         \\ 
 /  & / & RuVSb  & 0.44 &  & NiZrSn & 0.91 &  PdHfSn & 0.50 & 
 /  & / & NiScSb & 1.37 &  & \textbf{NiZrSn} & 2.01 &  FeNbSb & 1.65 &         \\       
/  & / & CoZrBi & 0.43 &  & RuVSb   & 0.88 &  NiYSb  & 0.49 & 
 / & / & IrZrSb & 1.36 &  & / &/&LiAsS$_{2}$& 1.65 &         \\ 
/  & / & NiYBi  & 0.43 &  & IrZrSb & 0.88 &   NiYBi  & 0.48 & 
 /  & / & \textbf{FeNbSb}& 1.36 &  & / & / &  \textbf{NiZrSn} & 1.62 &         \\ 
 /  & / & PdScSb & 0.43 &  & PdZrSn & 0.87 &  NiZrSn & 0.47 & 
 / & / & / & / &  & / & / &  / & / &         \\        
 /  & / & PtScSb & 0.42 &  & PdYSb &  0.87 &  IrZrSb & 0.47 & 
 /  & / & / & / &  & / & / &  / & / &         \\ 
 /  & /& \textbf{FeNbSb} & 0.41 &  & NiZrPb & 0.85 &  LiAsS$_{2}$ & 0.46 & 
 /  & / & / & / &  & / & / &  / & / &         \\        
 /  & / & / & / &  & PtYSb & 0.79 &  PdZrSn & 0.46 & 
 /  & / & / & / &  & / & / &  / & / &         \\        
 /  & / & / & / &  & \textbf{Fe$_{2}$VAl} & 0.79 &  NiZrPb & 0.45 & 
 /  & / & / & / &  & / & / &  / & / &         \\        
 /  & / & / & / &  & /& / &  AuScSn & 0.45 & 
 /  & / & / & / &  & / & /& /& / &         \\        
 /  & / & / & / &  & / & / &  PdLaBi & 0.44 & 
 /  & / & / & / &  & / & /  & / & / &         \\        
 /  & / & / & / &  & / & / &  NiScSb & 0.43 & 
 /  & / & / & / &  & / & / &  / & / &         \\        
 /  & / & / & / &  & / & / &  NiScBi & 0.42 & 
 /  & / & / & / &  & / & / &  / & / &         \\        
 /  & / & / & / &  & / & / &  PdScSb & 0.42 & 
 /  & / & / & / &  & / & / &  / & / &         \\        
 /  & / & / & / &  & / & / &  \textbf{Fe$_{2}$VAl} & 0.42 & 
 /  & / & / & / &  & / & / &  / & / &         \\

\hline
\hline
\end{tabular}
}
\end{table*}
\end{center} 

\section{Results and Discussion}

\begin{figure*}[t]
\centering
\includegraphics*[height=12cm,keepaspectratio]{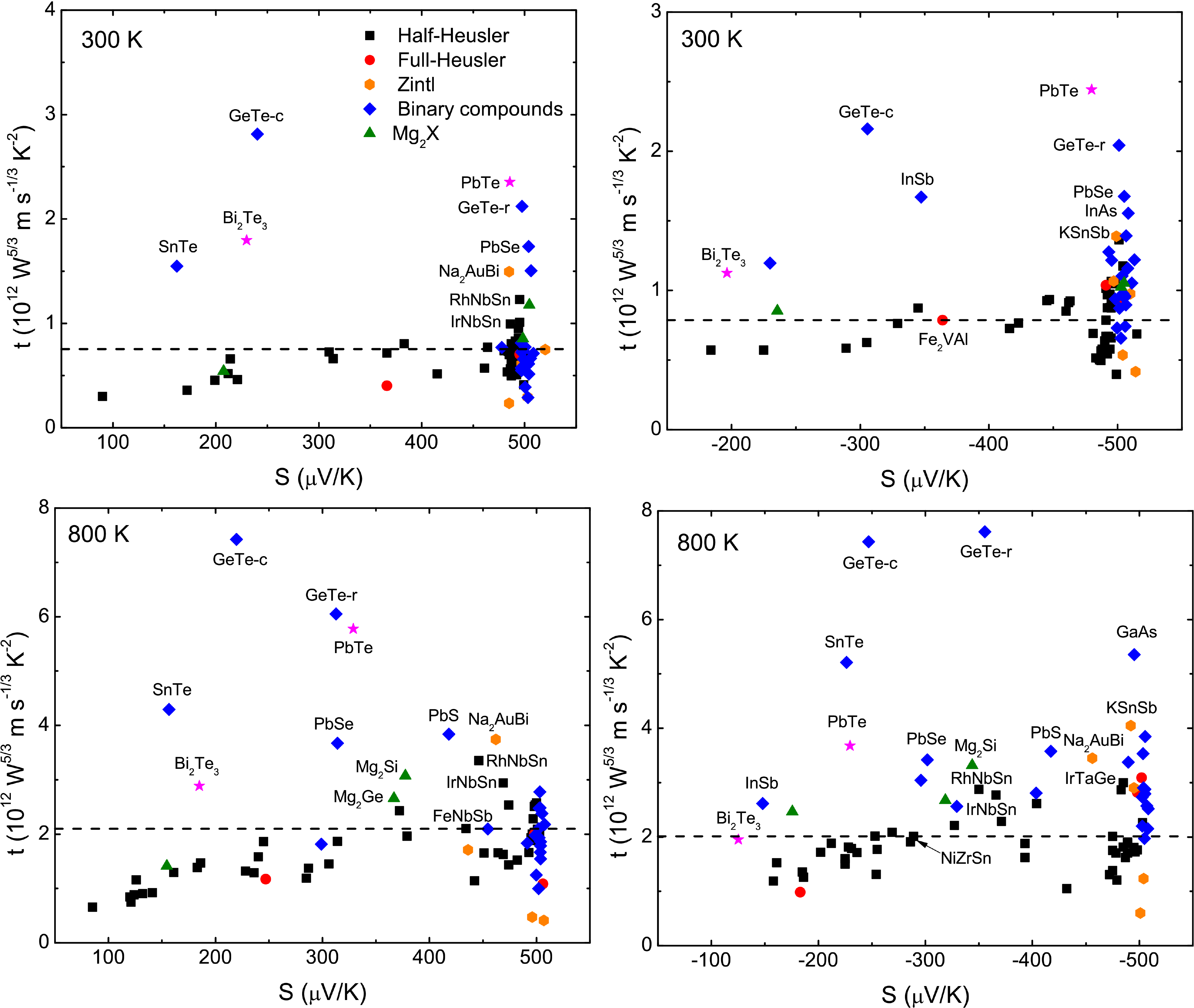}
\caption{\label{figTmaxS} The optimum EFF ($t$) and corresponding Seebeck coefficients at 300 and 800 K. The doted lines denote a screening criteria based on several known TE materials (see text).}
\end{figure*}

\begin{figure*}[t]
\centering
\includegraphics*[height=12cm,keepaspectratio]{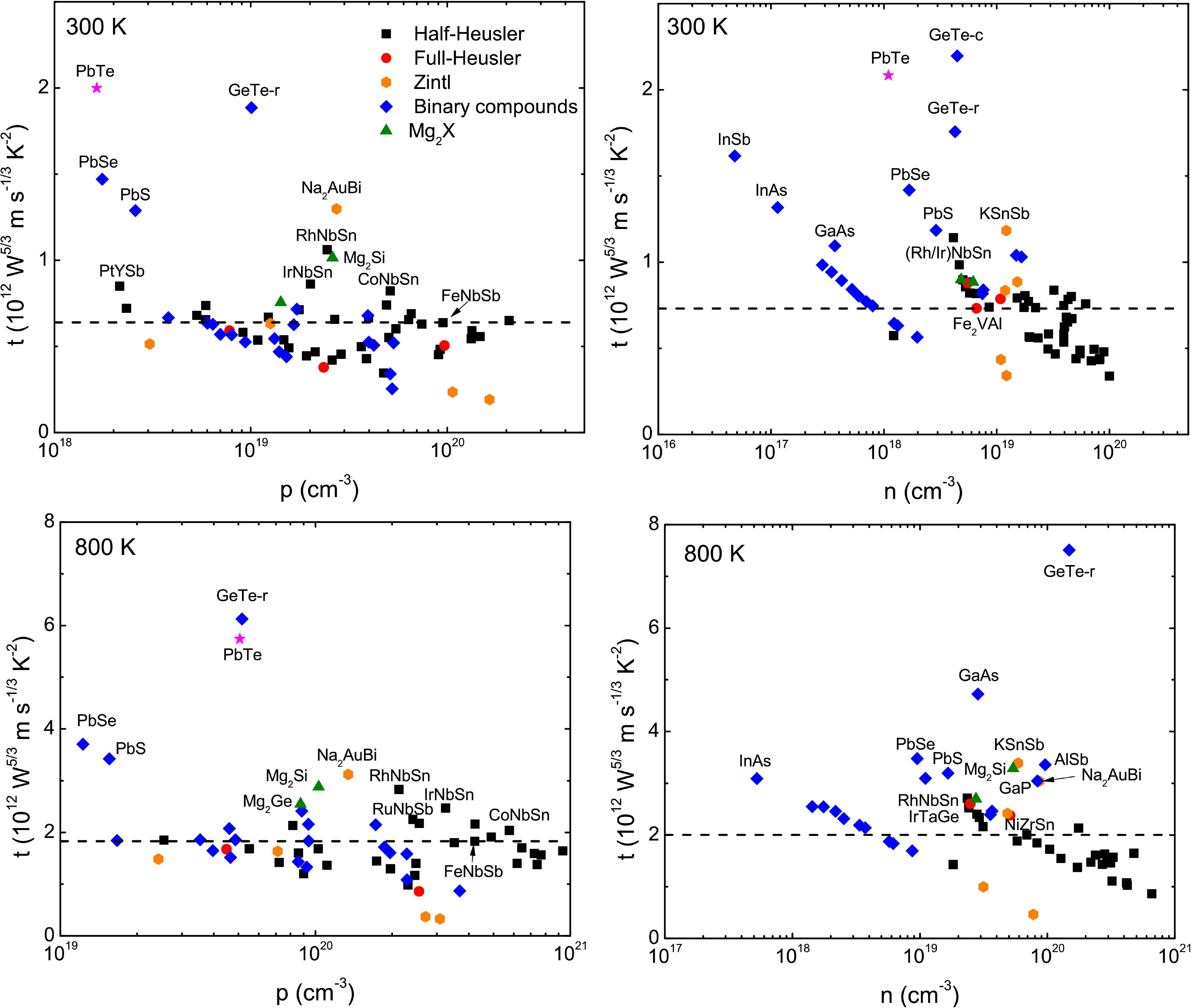}
\caption{\label{figTS300} EFF ($t$) and corresponding carrier concentrations at $S$ = 300 $\mu$V/K. The doted lines denote a screening criteria based on several known TE materials (see text). }
\end{figure*}

\begin{figure}[ t! ]
\centerline{\includegraphics[width=8cm]{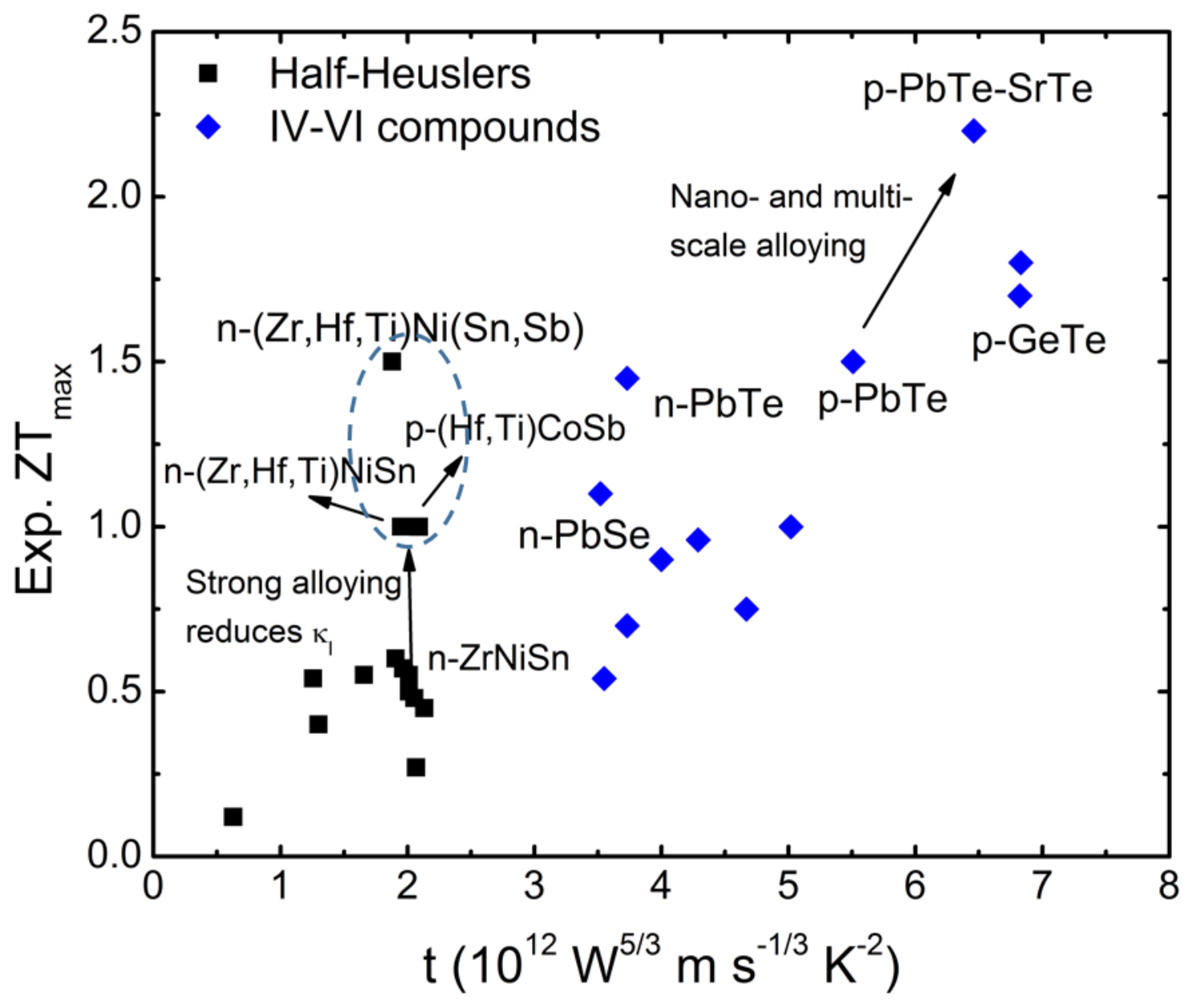}}
\caption{Experimental maximum $ZT$ for half-Heuslers and IV-VI semiconductors. These include GeTe,\cite{n3,ztGeTe1} PbTe,\cite{PbTe2,ztPbTe1,ztPbTe2} PbSe,\cite{ztPbSe1,ztPbSe2} PbS,\cite{ztPbS1,ztPbS2} SnTe,\cite{ztSnTe1,ztSnTe2,ztSnTe3} NiZrPb,\cite{ztNiZrPb} NiZrSn,\cite{ztNiZrSn1,ztNiZrSn2} NiTiSn,\cite{ztNiTiSn1,ztNiTiSn2} NiHfSn,\cite{ztNiZrSn1} PtHfSn,\cite{ztPtHfSn} CoZrSb,\cite{ztCoZrSb} CoNbSn,\cite{ztCoNbSn} PtYSb,\cite{ztPtYSb} NiYBi,\cite{ztNiYBi} and three strong-alloyed samples: (Zr,Hf,Ti)NiSn,\cite{ztTiZrHfNiSn} (Hf,Ti)CoSb,\cite{ztHfTiCoSb} and (Zr,Hf,Ti)Ni(Sn,Sb).\cite{ztTiZrHfNiSnSb} For those three strongly alloyed systems, the $t$ functions were calculated using pure NiZrSn, CoHfSb and NiTiSn, respectively. Note that the $t$ functions are different for the two $p$-PbTe points due to the different temperatures for the experimental maximum $ZT$ ($t$ is temperature dependent).}
\label{fig5}
\end{figure}

We indicate the screening criteria with the dotted lines in Figs. \ref{fig1} to \ref{figTS300}.
The compounds above the dotted line have EFF higher than the
reference compounds,
for which we choose the half-Heusler FeNbSb for $p$-type materials
at both 300 and 800 K.
This material has a high PF of 10.6$\times$10$^{-3}$ Wm$^{-1}$K$^{-2}$ at room temperature with Ti doping \cite{r28}
and optimum PFs from 4.3 to 5.5$\times$10$^{-3}$ Wm$^{-1}$K$^{-2}$ at 800 K.
\cite{r29}
For $n$-type materials, we used the full-Heusler Fe$_{2}$VAl for 300 K
and the half-Heulser NiZrSn for 800 K,
which show PFs of 5$\times$10$^{-3}$ Wm$^{-1}$K$^{-2}$
and 2-3$\times$10$^{-3}$ Wm$^{-1}$K$^{-2}$ at these temperatures, respectively.
\cite{r30,r31,r32}
These reference lines simply show the EFF for a good thermoelectric in order
to indicate a value for the EFF where it becomes interesting to study
materials in more detail.

As mentioned,
the EFF ($t$) identifies the complexity of the electronic structures in relation to TE performance, without regard to the origin of this complexity, as long
as it decouples $\sigma$ and $S$.
In the following we explore various ways of using it to identify promising materials. Note that some of the materials are anisotropic systems, and the EFF here is based on the directional-averaged properties, although it would be straightforward to apply it anisotropically using the anisotropic transport coefficients if such a screen was desired.

Fig. \ref{fig1} shows a scatter plot for peak values of the $t$ function and the corresponding doping levels at 300 and 800 K for both $p$- and $n$-type materials studied. Several of the high EFF materials are IV-VI binary semiconductors. At 800 K, $n$-type PbTe is inferior but GeTe still shows outstanding band EFF. Note that GeTe has a phase transition at 670 K and transforms from rhombohedral (GeTe-r) to cubic structure (GeTe-c).
We also find a fairly high $t$ value for $n$-GaAs,
but bulk GaAs is a known poor TE material due to the high thermal conductivity ($\sim$50 Wm$^{-1}$K$^{-1}$ at 300 K); thus we do not discuss it further. The detailed values of EFF and corresponding doping levels are summarized in Tables \ref{tab:table1} and S2. In terms of the ternary materials, the best materials are the little-studied Zintl compounds, KSnSb for $n$-type, and Na$_{2}$AuBi for both $p$- and $n$-type. The best several HHs have composition XYZ (X = Co, Rh, Ir; Y = Nb, Ta; Z = Ge, Sn), and then some antimonides as shown in Table. \ref{tab:table1}. Among these HH candidates, CoNbSn has recently been reported with an enhanced $n$-type $ZT \sim$ 0.6.\cite{r33} Another recent theoretical study also predicted Co(Nb,Ta)Sn as potential promising $p$-type TE materials.\cite{r14} In addition to the reference compounds, some other known good TE materials such as $p$-type CoHfSb and $n$-type NiHfSn are included. $n$-type Mg$_{2}$X show more favorable band structure than the $p$-type counterparts. However, Mg$_{2}$Si and Mg$_{2}$Ge also have reasonably complex $p$-type electronic structures as seen in the EFF. Moreover, the little studied FH compounds Li$_{2}$NaSb and K$_{2}$CsSb show larger $n$-type EFF than Fe$_{2}$VAl, especially at 800 K.

For a degenerate doped single parabolic band, the Seebeck coefficient (low $T$) is given by
$S(T,n)$ = $\frac{8\pi^{2}k^{2}_{B}T}{3eh^{2}}m^{\ast}_{\mathrm{DOS}}$($\frac{\pi}{3n}$)$^{2/3}$.
In a more general case at low $T$, $S$ maintains an inverse relationship
with $E_F$, $S$ $\propto$ 1/$E_F$,
where $S$ is independent of $m^{\ast}$ at fixed Fermi energy (here $E_F$
is relative to the band edge).
This suggessts that at fixed $E_F$ and temperature,
$S$ should be similar for different materials.
Certain band features such as flat bands near $E_F$ can enhance $S$ by increasing the energy dependence of the conductivity, as in $n$ type lanthanum telluride,\cite{vr3} where a heavy band near the light band extrema enhances $S$. Thus one can imagine another approach focusing on $t$ at a fixed energy, rather than the peak value. Fig. \ref{fig2} presents the iso-energy EFF at 0.05 eV away the band edges. As can be seen, the trend is similar to that in Fig. \ref{fig1}, but with a higher carrier concentration (10$^{19}$ $\sim$ 10$^{21}$ cm$^{-3}$), and less variation among the high $t$ materials.

It is also informative to examine the utility of the electronic fitness function in relation to the Seebeck coefficient. Good bulk TE materials usually have $S \sim$ 200 - 300 $\mu$V/K. In fact from the Wiedemann-Franz relation, the electronic contribution of thermal conductivity can be formulated as $\kappa_{e}$ = $L\sigma T$. $L$ = 2.45$\times$10$^{-8}$ W$\Omega$/K$^{2}$ is the standard Lorenz number. Then one can rewrite $ZT$ as $ZT = rS^{2}/L$ where $r = \kappa_{e}/(\kappa_{e}+\kappa_{l})$, $\kappa_{e}$ and $\kappa_{l}$ are the electronic and lattice thermal conductivity. Even assuming the extreme case with $r$ = 1 which means $\kappa_{l}$ = 0, $S$ has to be larger than 156 $\mu$V/K to achieve $ZT$ = 1. 

We show the peak $t$ values and the corresponding Seebeck coefficients in Fig. \ref{figTmaxS}. As can be seen, $S$ mainly fall into around 500 $\mu$V/K at 300 K for both $p$- and $n$-type, especially for the binary compounds. But at 800 K, several known high performance IV-VI tellurides exhibit $S$ $\sim$200-300 $\mu$V/K. On the other hand, those best HHs, FHs and Zintls all show larger $S$ $\sim$350-500 $\mu$V/K for both $p$- and $n$-type cases. Finding high Seebeck but with reasonable doping is critical since large $S$ in low $\sigma$, low PF compounds usually corresponds to low doping where the lattice thermal conductivity dominates and leads to low performance. On the other hand at high doping levels one will find small $S$ which is unfavorable even with high $\sigma$. A balance is needed. Therefore in Fig. \ref{figTS300} we show the EFF and corresponding doping levels at $S$ = 300 $\mu$V/K. One can see that the best materials as indicated in the maximum and isoenergy $t$ functions (Figs. \ref{fig1} and \ref{fig2}) still show larger $t$ values compared to other materials. Note that some materials that do not possess such high $S$ at any doping levels are not shown. 

The efficacy of the electronic fitness function can be assessed from comparison with existing experimental data. In Fig. 5 we present the maximum experimental $ZT$ at different temperatures and corresponding optimum $t$ functions. In addition to the electronic structure complexity, two other ingredients can affect $ZT$, which are the scattering mechanism and lattice thermal conductivity. We show experimental data for compounds and nearby alloys. We also selected several heavily isoelectronic-alloyed samples for comparison. We focus on the HHs and the IV-VI semiconductors due to the availability of the experimental data. From Fig. 5, one can see that the optimized experimental $ZT$ increases with increasing EFF, especially for the binary semiconductors. $ZT$ can be enhanced due to the reduction of $\kappa_l$ as in isoelectronic alloying in the HHs and by multi-scale approach in $p$-PbTe.\cite{isoelectronic1,isoelectronic2,ztPbTe2} In the following, we will focus on the several best candidates in binary and ternary compounds and explicitly discuss the details of the electronic structures to explore the superior features for potential TE performance.

\begin{figure}[ t! ]
\centerline{\includegraphics[width=8.5cm]{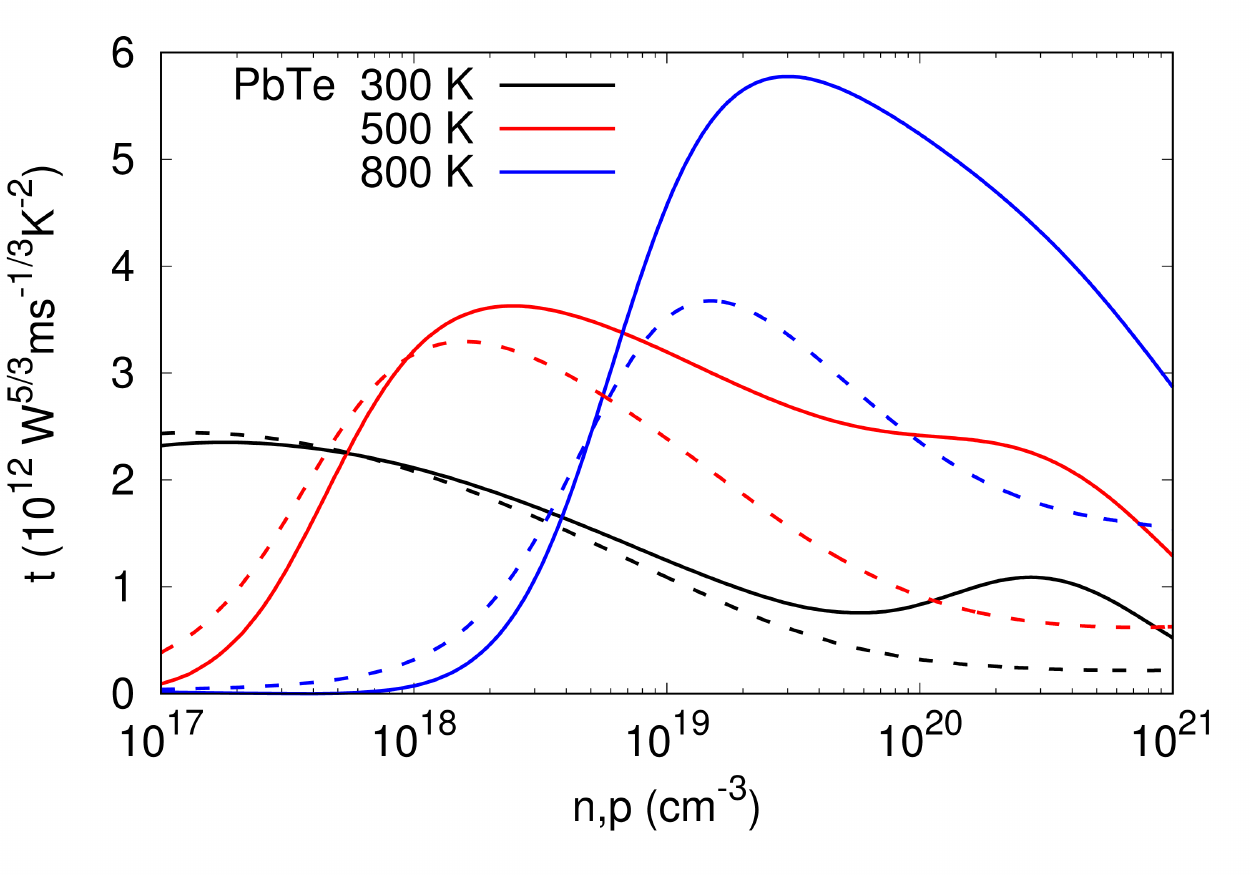}}
\caption{EFF ($t$) with respect to the carrier concentration at different temperatures for PbTe. Solid lines and dash lines represent $p$- and $n$-type compounds, respectively}
\label{PbTe-T}
\end{figure}

\begin{figure}[ t! ]
\centerline{\includegraphics[width=8.5cm]{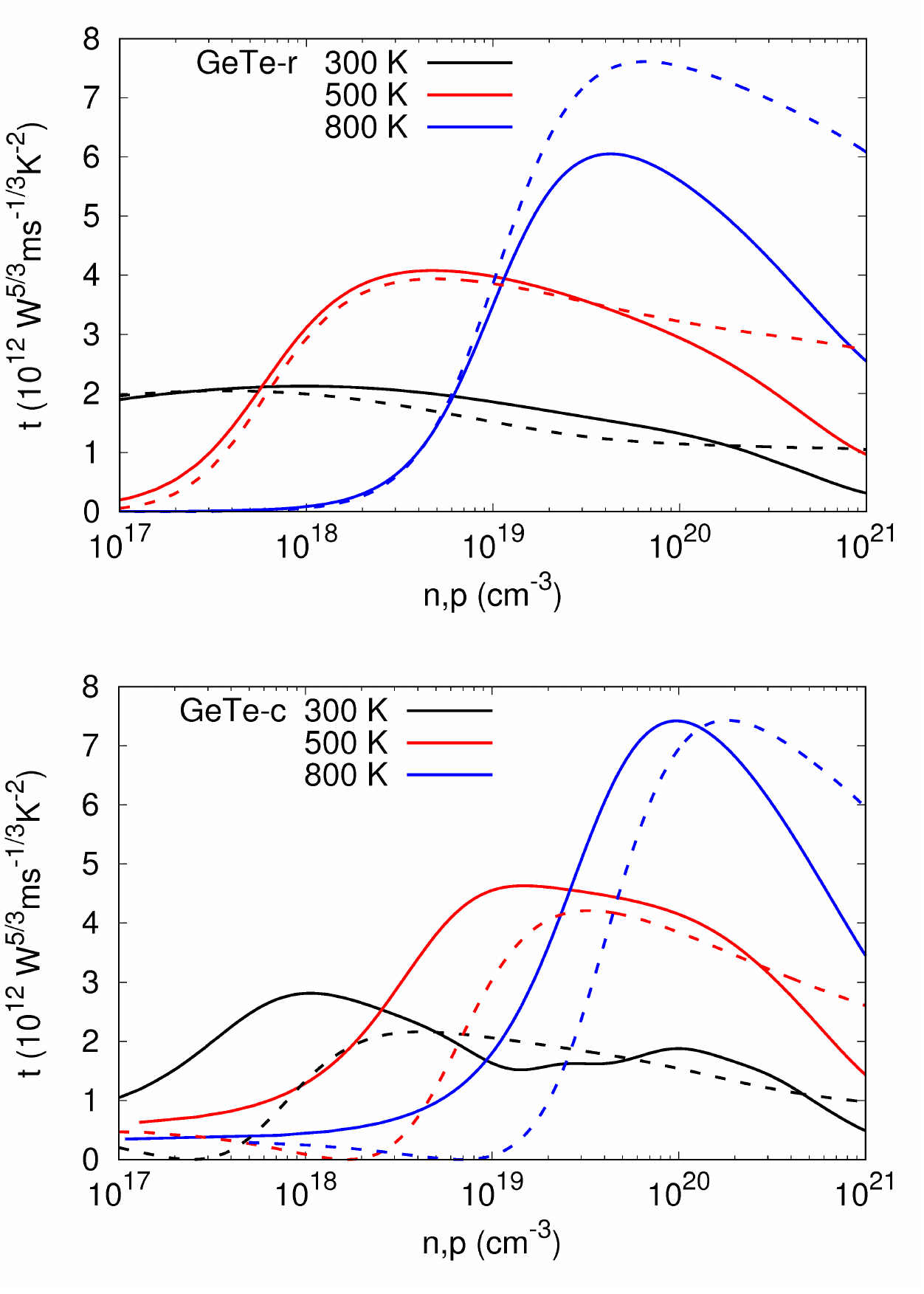}}
\caption{EFF ($t$) with respect to the carrier concentration at different temperatures for rhobohedral GeTe and cubic GeTe. Solid lines and dash lines represent $p$- and $n$-type compounds, respectively}
\label{GeTe-T}
\end{figure}

\begin{figure}[t!]
\centerline{\includegraphics[width=8.5cm]{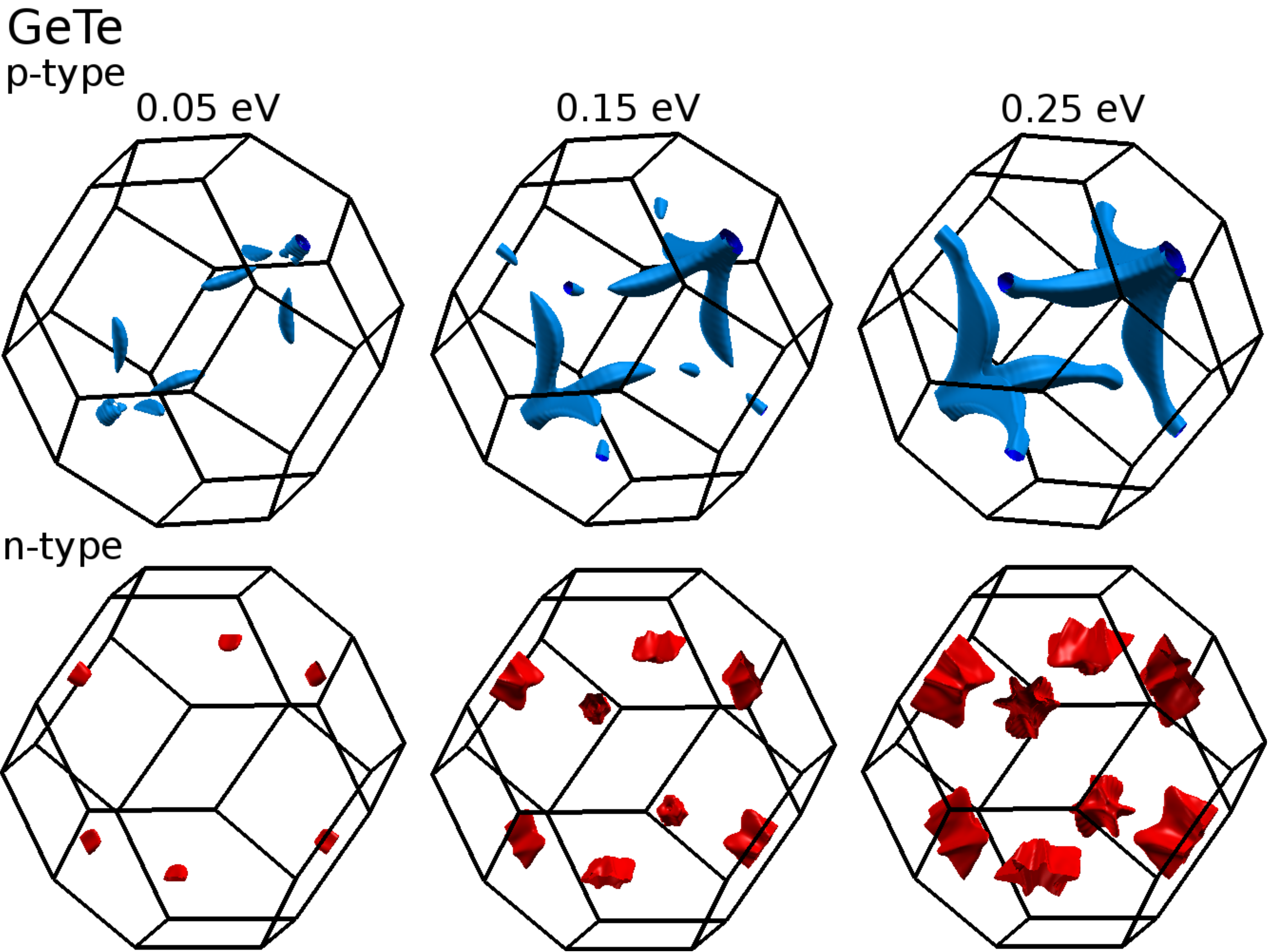}}
\caption{Calculated constant energy surfaces of rhobohedral GeTe. Iso-energies at 0.05 eV, 0.15 eV and 0.25 eV below the VBM ($p$-type with blue color) and above the CBM ($n$-type with red color are depicted.}
\label{GeTe}
\end{figure}

\begin{figure*}[t]
\centering
\includegraphics*[height=10cm,keepaspectratio]{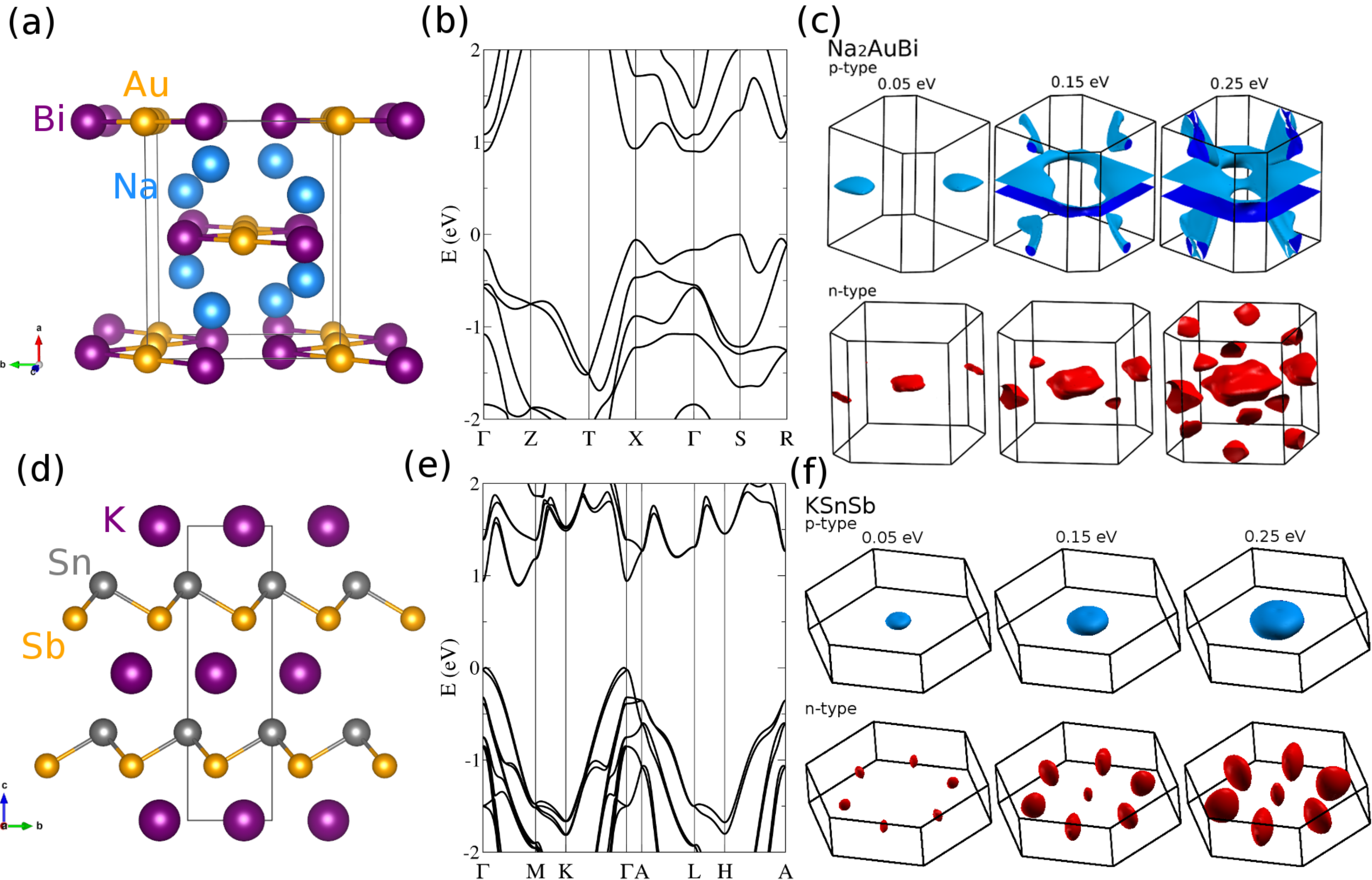}
\caption{\label{zintl} The crystal structures, band structures and isosurfaces of Na$_{2}$AuBi ((a)-(c)) and KSnSb ((d)-(f)).}
\end{figure*} 


PbTe is a state-of-art TE material for the middle-to-high temperature range ($\sim$ 800 K). The EFF (Fig. \ref{PbTe-T}) is high due to the valley degeneracy at L point and the secondary band contribution along the $\Sigma$ line, which can be clearly observed in the iso-surface plots (Fig. S1) where a connected surface is presented at heavily $p$-type doped case ($\sim$ 0.25 eV). The energy difference between L ($N_v$ = 4) and $\Sigma$ ($N_v$ = 12) bands are found to be reduced with elevated temperatures.\cite{isoene3} The EFF shows (Fig. \ref{PbTe-T}) better $p$-type performance as temperature increases.

GeTe is relatively less studied. Recent studies show promising TE performance of $p$-type GeTe with a $ZT$ peak value of 1.8.\cite{n3} The EFF of cubic and rhombohedral structures are shown in Fig. \ref{GeTe-T}. Both structures show larger $n$-type EFF values at higher doping levels. This is due to the secondary conduction band contribution in both materials. The superior performance in both $p$- and $n$-type GeTe can be attributed to the high band degeneracy and complex iso-energy surfaces, as observed from the iso-surfaces in Fig. \ref{GeTe} and Fig. S2. As seen, $n$-type GeTe has degenerate valleys at L points with corrugated shapes in both structure types. On the other hand, $p$-type GeTe-c has the VBM also at L point and with a nearby valence band along the $\Sigma$ line that will contribute when heavily doped, which is clearly shown in the isosurface similar to cubic PbTe. GeTe-r shows a dominant $\Sigma$ band at VBM with six valleys inside the Brillouin zone at low doping. These hole pockets become connected with the L-point-valley at higher doping levels.




Zintl phases have been explored and reported as promising TE materials.\cite{r44,r45} Here we choose several representatives which were shown to have potential promising TE performance theoretically.\cite{r15} Based on the EFF, two alkali metal Zintl phases Na$_{2}$AuBi and KSnSb are identified to have particularly favorable band complexity compared with the other ternary compounds including selected HHs and FHs. 

Na$_{2}$AuBi has been identified as a semiconductor theoretically,\cite{r46} but is little-studied as a TE material. It crystallizes with an orthorhombic $Cmcm$ structure (Fig. \ref{zintl} (a)) which can be viewed as poly-anionic [AuBi]$^{2-}$ layers separated by the Na cations along the $a$ axis. It is worth noting that the [AuBi]$^{2-}$ layers form a 1-D zigzag ``ribbon''. Na$_{2}$AuBi has an indirect band gap with VBM at S and CBM along $\Gamma$-S, as shown in Fig. \ref{zintl} (b). The most striking feature is the multiple band extrema near the band edges ($<$ 0.25 eV) for both $p$- and $n$-type materials. Moreover, one may notice a combination of heavy and light bands at both VBM and CBM. These features are favorable for achieving high Seebeck and conductivity reflected in a high EFF. We present the iso-surface plots in Fig. \ref{zintl} (c). As seen, the isosurfaces have complex shapes.  For $p$-type, the hole pockets are very anisotropic at low doping and become low-dimensional sheet-like surfaces as doping increases. Four more low-dimensional pockets around R point are also seen at higher doping levels. In $n$-type case, anisotropic pockets are also seen at $\Gamma$ and X points at 0.05 eV. These pockets develop to more corrugated surfaces with other pockets at higher doping levels. It is clear that both $p$- and $n$-type Na$_{2}$AuBi show large band degeneracy and substantial low-dimensional anisotropic carrier pockets, which are favorable for high TE performance. This is reflected in the doping-dependent EFF plot (Fig. S3), where both $p$- and $n$-type Na$_{2}$AuBi show large $t$ values with $p$-type more favorable.

KSnSb has received attention recently as a promising TE material. The $n$-type material has been theoretically shown to have high mobility and large band degeneracy.\cite{r47} KSnSb adopts a hexagonal crystal structure with anionic [SnSb]$^{-}$ layers stacking along the $c$-direction and separated by K$^{+}$ slabs (Fig. \ref{zintl} (d)). This structure motif is quite similar to the 122 phases and Mg$_{3}$Sb$_{2}$, for which excellent $n$-type performance was predicted theoretically and found recently by experiment in Mg$_{3}$Sb$_{2}$.\cite{r5,r48} Similarly, we find that KSnSb also shows higher $n$-type EFF (Fig. \ref{fig1} - \ref{figTS300}). From the band structure (Fig. \ref{zintl} (e)), the CBM is along $\Gamma$-M direction with next conduction band extrema at only 51 meV higher in energy. Moreover, the bands along the in-plane direction ($\Gamma$-M and $\Gamma$-K ) are seen to be more dispersive than out-of-plane direction ($\Gamma$-A). All these yield higher band degeneracy and larger conductivity in the in-plane direction, which can be visualized in the iso-energy plots (Fig. \ref{zintl} (f)). Notice that those electron pockets are inside the Brilloin zone, which yield a valley degeneracy of 7 (including $\Gamma$). They also exhibit anisotropic character, especially at zone center. On the other hand, $p$-type KSnSb shows spherical shapes of pockets at zone center, which is inferior to the $n$-type counterpart, as clearly shown in the doping-dependent $t$ function plot (Fig. S3). Note also that the reaction temperatures for synthesis of Na$_{2}$AuBi and KSnSb are 973 K and 803 K,\cite{add8,add9} respectively.


\begin{figure}[t]
\centering
\includegraphics*[height=3.4cm,keepaspectratio]{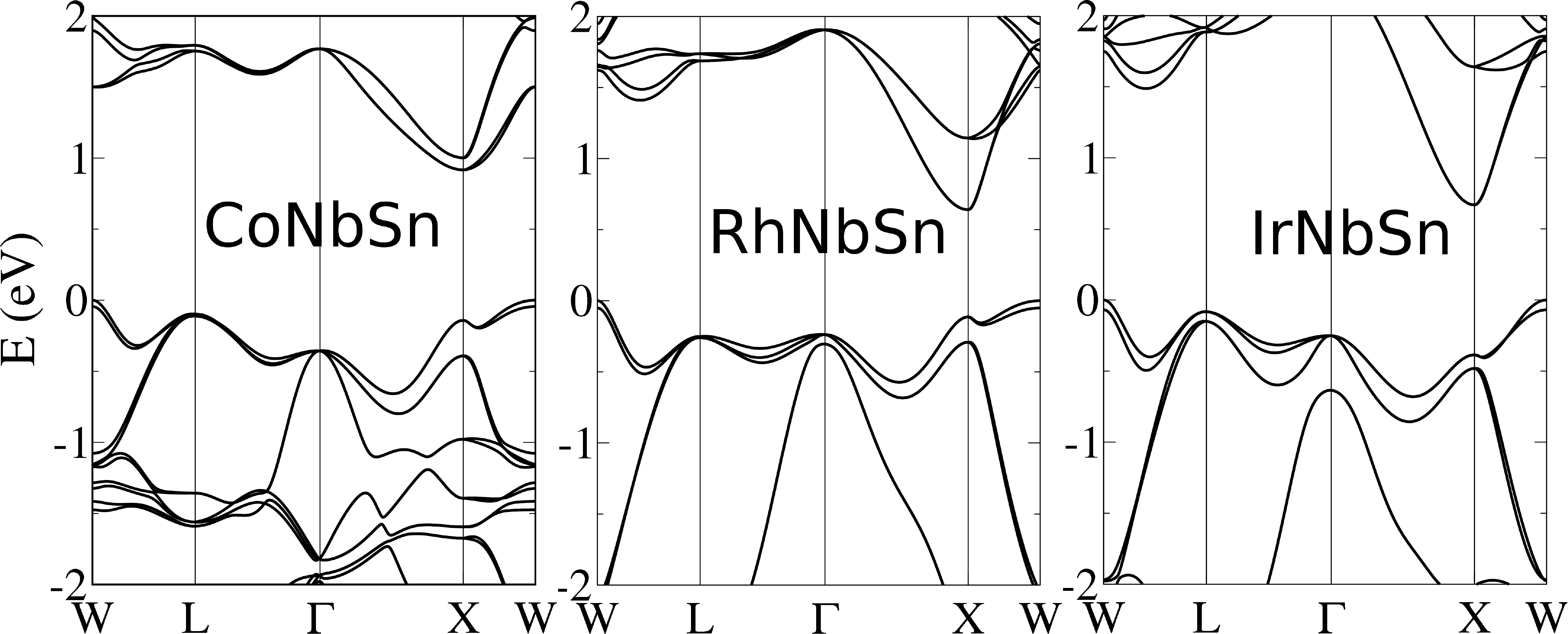}
\caption{\label{fig11} Calculated band structures of XNbSn (X = Co, Rh, Ir) compounds.}
\end{figure}

\begin{figure}[h!]
\centering
\includegraphics*[height=18cm,keepaspectratio]{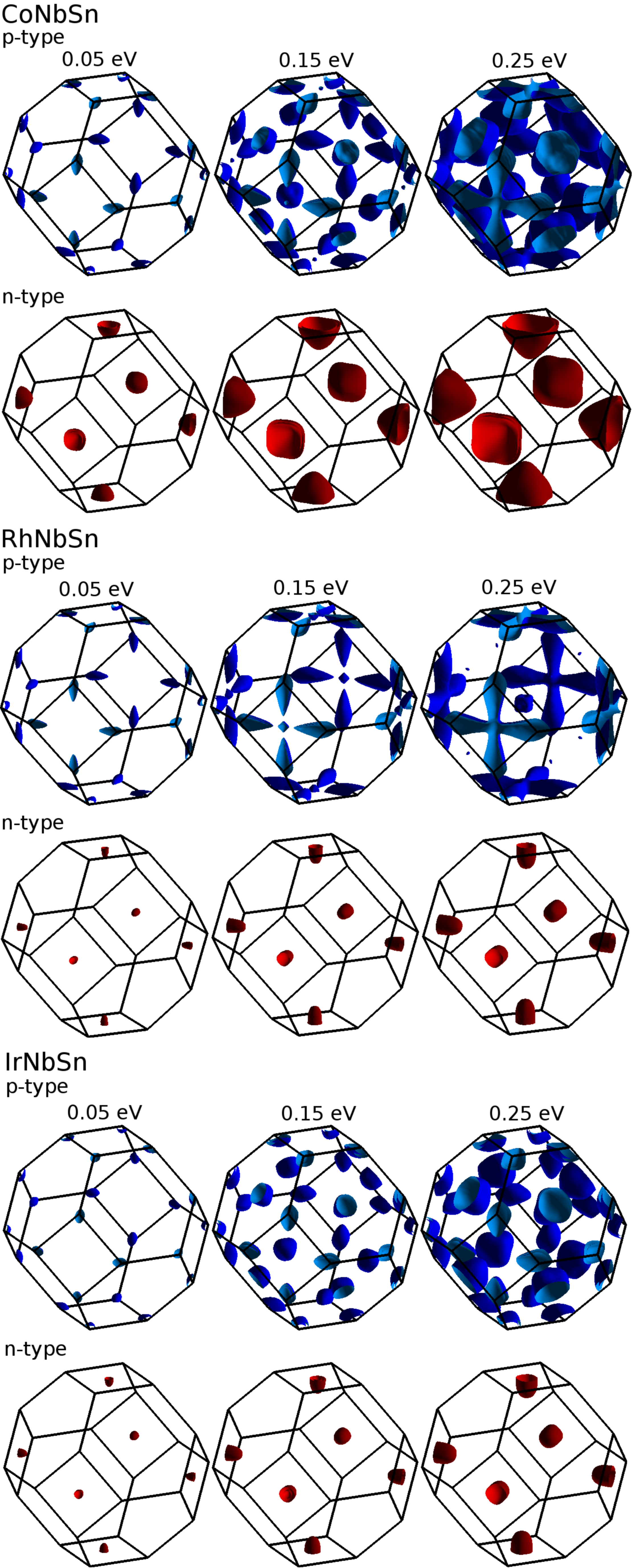}
\caption{\label{fig12} Calculated constant energy surfaces of XNbSn (X = Co, Rh, Ir). Iso-energies at 0.05 eV, 0.15 eV and 0.25 eV below the VBM ($p$-type with blue color) and above the CBM ($n$-type with red color) are depicted.}
\end{figure}

\begin{figure}[h!]
\centering
\includegraphics*[height=15cm,keepaspectratio]{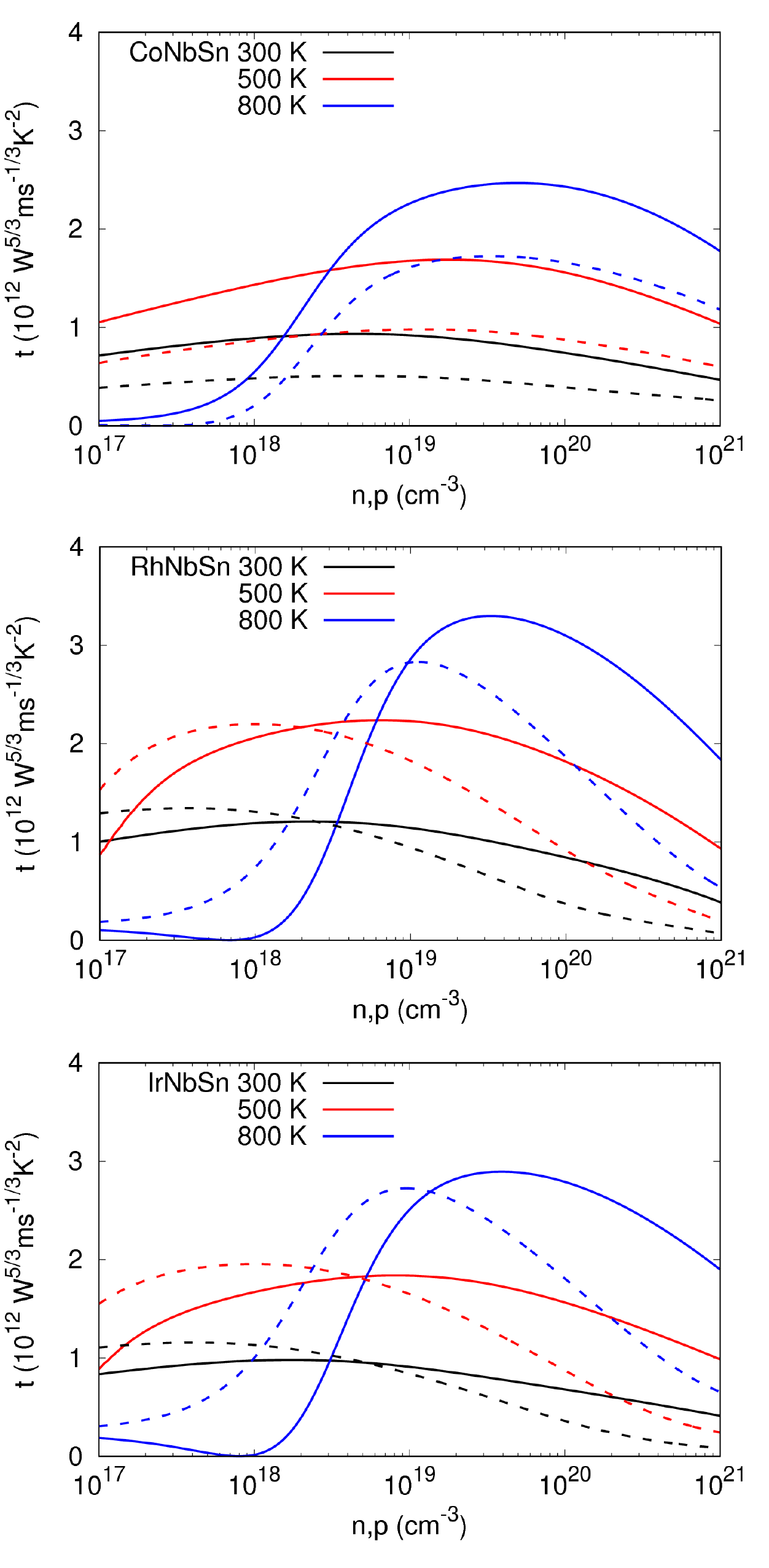}
\caption{\label{fig13} EFF ($t$) with respect to the carrier concentration at different temperatures for XNbSn (X = Co, Rh, Ir). Solid lines and dashed lines represent $p$- and $n$-type compounds, respectively.}
\end{figure} 

\begin{figure}[h!]
\centering
\includegraphics*[height=12cm,keepaspectratio]{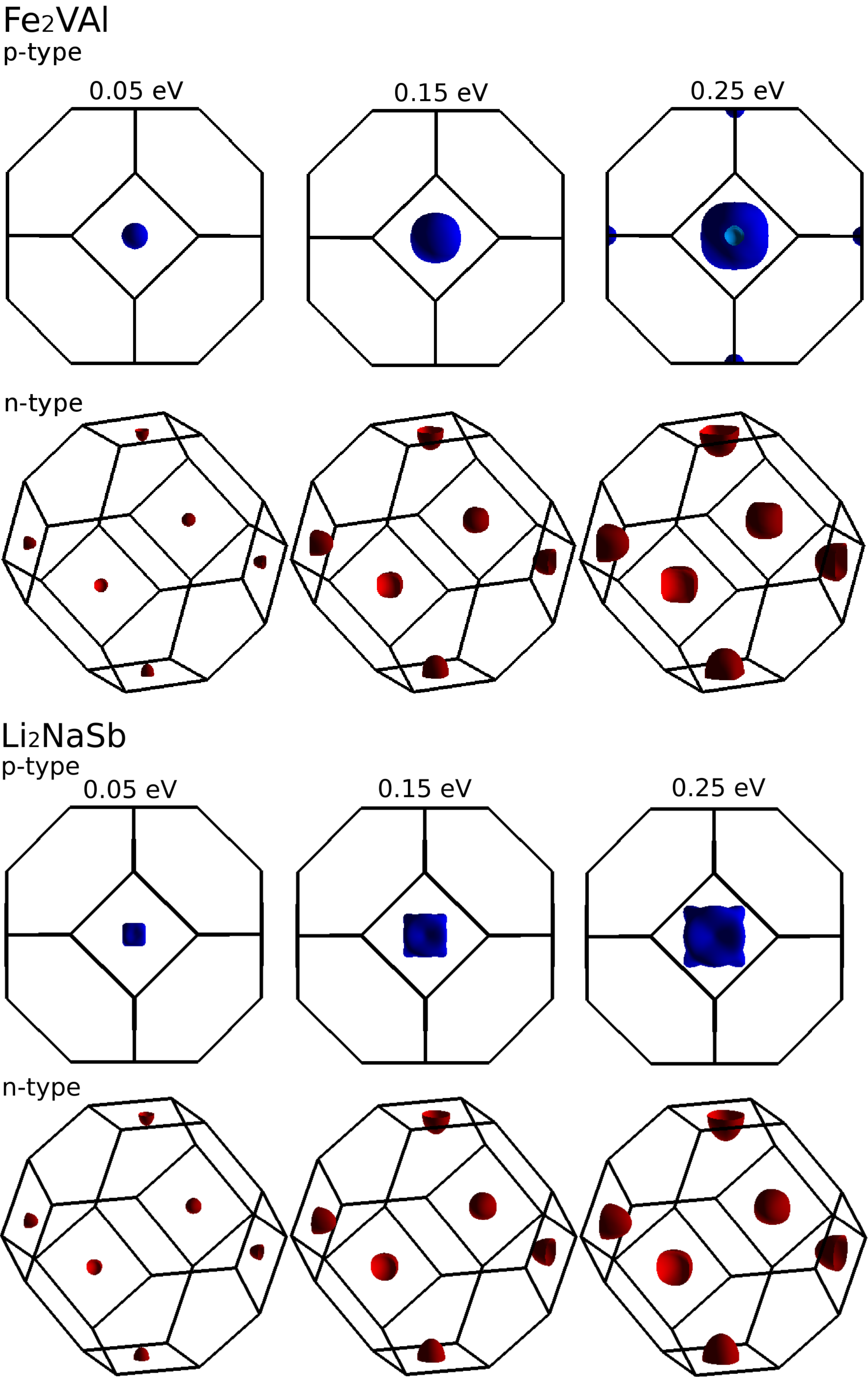}
\caption{\label{fig14} Calculated constant energy surfaces of Fe$_{2}$VAl and Li$_{2}$NaSb. Iso-energies at 0.05 eV, 0.15 eV and 0.25 eV below the VBM ($p$-type with blue color) and above the CBM ($n$-type with red color) are depicted.}
\end{figure} 

Heusler compounds are intermetallics with a face-centered
cubic structure.
They are further divided into full- and half-Heuslers
based on the occupation of the body diagonal positions. 
The HH phases studied here are based on the experimentally known MgAgAs structure-type and with a 18 valence electrons count, which include 42 semiconducting compounds. The HHs have been explored widely as potential TE materials.\cite{r11,add3,add4,add5} As seen in Table \ref{tab:table1}, the best $p$-type HHs mainly fall into the XNbSn (X = Co, Rh, Ir) compounds at both 300 and 800 K. On the other hand, RhNbSn, IrNbSn and IrTaGe are the best $n$-type at both temperatures. However for real applications, the Ir- and Rh-compounds have limitations due to the cost of these elements. Nevertheless, we focus on the XNbSn (X = Co, Rh, Ir) compounds to discuss the electronic features that underlie the high EFF. The band structures are shown in Fig. \ref{fig11}. Based on the crystal symmetry, the carrier pockets at W have higher band degeneracy than X point. Moreover, while details differ near the VBM where they all show multiple band extrema close in energy ($\sim$0.25 eV). This can be seen in the constant energy surfaces as shown in Fig. \ref{fig12}, where the hole pockets also show anisotropy. The higher band degeneracy, valley anisotropy and multi-band contributions at deeper energy in valence bands yield larger EFF in $p$-type compounds especially at high doping levels (Fig. \ref{fig13}).

In contrast to the HH compounds, transition-metal containing semiconducting FHs are vary rare due to the Slater-Pauling behavior. \cite{r37} Nonetheless, FHs with 24 valence electrons per formula unit can be semiconductors.\cite{r38,add2} Fe$_{2}$VAl is such an example with a large PF at room temperature (4-6 mW m$^{-1}$K$^{-2}$ \cite{r30,r39,r40}), though the $ZT$ is only around 0.13-0.2 due to the high thermal conductivity.\cite{r39,r41}
Li$_{2}$NaSb and K$_{2}$CsSb exemplify another potential semiconducting FH
system that does not include transition elements and has received
less attention.\cite{r42}
Here we focus on the semiconducting FHs,
which include two non-transition-metal compounds.
We find that both Li$_{2}$NaSb and K$_{2}$CsSb show
very promising $n$-type EFF (Table \ref{tab:table1}). K$_{2}$CsSb has been known as an excellent photocathode materials. This is a property that correlates with air sensitivity as it requires low electron affinity and thus ready oxidation. Therefore here we focus on the best one (Li$_{2}$NaSb) together with the known Fe$_{2}$VAl for comparison. Note that the reaction temperature of Li$_{2}$NaSb is 1133 K \cite{add10} which implies that this material might be stable at 800 K. 

We plot the constant energy surfaces of these two materials in Fig. \ref{fig14}. One finds similar electron pockets in both Fe$_{2}$VAl and Li$_{2}$NaSb. Both the $p$-type materials show pockets at $\Gamma$ point. However, at high doping levels, $p$-type Fe$_{2}$VAl also shows contributions from X point. This is also seen in the band structures (Fig. S4) where the X-point-band is about 0.19 eV lower than VBM in Fe$_{2}$VAl. This suggests potential better performance of $p$-type Fe$_{2}$VAl at high doping concentrations which is exactly seen in the EFF plot (Fig. S5). Furthermore, the energy surfaces are more corrugated at $\Gamma$ in $p$-type Li$_{2}$NaSb compared to Fe$_{2}$VAl. Hence at moderate-to-low doping levels, $p$-type Li$_{2}$NaSb exhibits larger $t$. Another feature is the effect of bipolar conduction in Fe$_{2}$VAl due to the smaller band gap. This bipolar effect is detrimental to the TE performance and is clearly seen in Fig. S5. The two materials show similar EFF at 300 K, while at high temperatures and even moderate doping levels (e.g., at 800 K), the EFF drops dramatically as carrier concentration decreases especially in Fe$_{2}$VAl.

\section{Summary and Conclusions}

Thus in all these varied cases with different types of complex electronic structures the electronic fitness function captures the behavior that is favorable for the thermoelectric performance.
When $t$ is high, the material is dopable to the needed level and $\kappa$ is low, high $ZT$ results. Therefore, we present this simple transport function that describes the electronic aspect of $ZT$. It is based on the first-principles electronic structure and Boltzmann transport theory and is easily calculated. The essential aspect of this function is that it is large for band structures that overcome the inverse relationship between $\sigma$ and $S$ through complex shapes, multi valleys, heavy-light mixtures, band convergence, valley anisotropy, and other features. By applying this function to a large library of 75 potential TE materials, we have demonstrated that this function can efficiently screen the materials. The electronic fitness function also predicts two promising alkali metal Zintl compounds, KSnSb for $n$-type and Na$_{2}$AuBi for $p$- and $n$-type at both 300 and 800 K. Importantly, we identified some novel $p$- and $n$-type promising TE HHs. We also identified two semiconducting full-Heuslers, Li$_{2}$NaSb and K$_{2}$CsSb, which may show better $n$-type performance compared to Fe$_{2}$VAl. The EFF provides a simple and easy to use method to screen materials for potential TE performance.

\section*{Acknowledgments}

Work at University of Missouri was supported by the Department of Energy through the S3TEC Energy Frontier Research Center award \# DE-SC0001299/DE-FG02-09ER46577. G.X. gratefully acknowledges support from the China Scholarship Council.

\bibliography{tfunction}

\end{document}